\documentclass[a4paper]{article}

\usepackage{INTERSPEECH2022}
\usepackage{multirow}
\usepackage{bm}
\usepackage{cite}
\usepackage{array}

\title{Paraformer: Fast and Accurate Parallel Transformer for Non-autoregressive End-to-End Speech Recognition}
\name{Zhifu Gao$^1$, Shiliang Zhang$^1$, Ian McLoughlin$^2$, Zhijie Yan$^1$}
\address{
  $^1$Speech Lab, Alibaba Group, China\\
  $^2$ICT Cluster, Singapore Institute of Technology, Singapore}
\email{\{zhifu.gzf, sly.zsl\}@alibaba-inc.com, ian.mcloughlin@singaporetech.edu.sg}

\begin{document}

\maketitle
\begin{abstract}
Transformers have recently dominated the ASR field. Although able to yield good performance, they involve an autoregressive (AR) decoder to generate tokens one by one, which is computationally inefficient. To speed up inference, non-autoregressive (NAR) methods, e.g. single-step NAR, were designed, to enable parallel generation. However, due to an independence assumption within the output tokens, performance of single-step NAR is inferior to that of AR models, especially with a large-scale corpus. There are two challenges to improving single-step NAR: Firstly to accurately predict the number of output tokens and extract hidden variables; secondly, to enhance modeling of interdependence between output tokens. To tackle both challenges, we propose a fast and accurate parallel transformer, termed \emph{Paraformer}. This utilizes a continuous integrate-and-fire based predictor to predict the number of tokens and generate hidden variables. A glancing language model (GLM) sampler then generates semantic embeddings to enhance the NAR decoder's ability to model context interdependence. Finally, we design a strategy to generate negative samples for minimum word error rate training to further improve performance. Experiments using the public AISHELL-1, AISHELL-2 benchmark, and an industrial-level 20,000 hour task demonstrate that the proposed \emph{Paraformer} can attain comparable performance to the state-of-the-art AR transformer, with more than 10x speedup.\footnote{
% We will release code soon. 
This work was supported in part by Key R \& D Projects of the Ministry of Science and Technology (2020YFC0832500)}.
\end{abstract}

\section{Introduction}
\label{sec:intro}

Over the past few years, the performance of end-to-end~(E2E) models has surpassed that of conventional hybrid systems on automatic speech recognition~(ASR) tasks.
There are three popular E2E approaches: connectionist temporal classification (CTC)~\cite{graves2006connectionist}, recurrent neural network transducer (RNN-T)~\cite{graves2013speech} and attention based encoder-decoder (AED)~\cite{chan2016listen,vaswani2017attention}. 
Of these, AED models have dominated seq2seq modeling for ASR, due to their superior recognition accuracy. Examples are Transformer~\cite{vaswani2017attention} and Conformer~\cite{gulati2020conformer}. 
While performance is good, the auto-regressive~(AR) decoder inside such AED models needs to generate tokens one by one, since each token is conditioned on all previous tokens.
Consequently, the decoder is computationally inefficient, and decoding time increases linearly with the output sequence length.
To improve efficiency and accelerate inference, non-autoregressive (NAR) models have been proposed to generate output sequences in parallel~\cite{gu2018non,lee2018deterministic,ghazvininejad2019mask}.

\begin{figure}[t]
	\centering
	\includegraphics[width=0.8\linewidth]{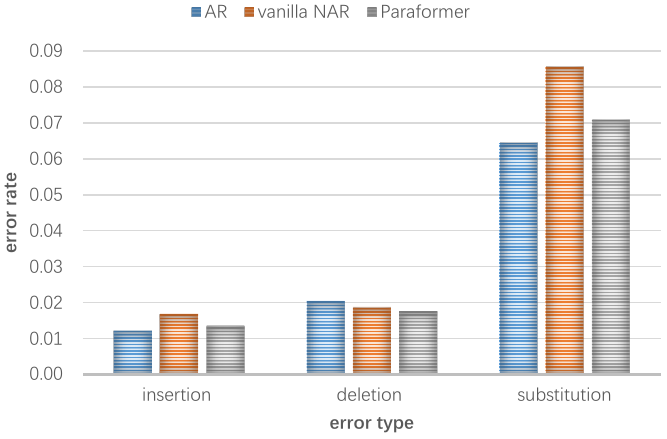}
	\caption{Analysis of different error types for three systems, evaluated on the industrial 20,000 hour task. }
	\label{fig:err}
	\vspace{-6mm}
\end{figure}
%

% Given the above, there is growing interest to exploit NAR models in the field of ASR. 
Based on the number of iterations duration inference, NAR models can be categorized as either \textit{iterative} or \textit{single step}.
Among the former, 
A-FMLM was the first attempt~\cite{chen2019listen}, designed to predict masked tokens conditioned on unmasked ones over constant iterations.
Performance suffers due to the need to predefine a target token length.
\
To address this issue, Mask-CTC and variants proposed enhancing the decoder inputs with CTC decodings~\cite{higuchi2020mask,higuchi2021improved,song2021non}.
\
Even so, these iterative NAR models require multiple iterations to obtain a competitive result, limiting the inference speed  in practice.
\
More recently, several single step NAR models were proposed to overcome this limitation~\cite{tian2020spike,fan2021cass,fan2021improved,chen2021align,deng2022improving}.
These generate output sequences simultaneously by removing temporal dependency.
\
Although single step NAR models can significantly improve inference speed, their recognition accuracy is significantly inferior to AR models, especially when evaluated on a large-scale corpus.
\

The single step NAR works mentioned above mainly focus on how to predict token numbers as well as extract hidden variables accurately. 
Compared to machine translation which predicts token number by a predictor net, it is indeed difficult for ASR due to various factors such as the speaker's speech rate, silences, and noise.
On the other hand, according to our investigation, single step NAR models make a lot of substitution mistakes compared to AR models~(Depicted as AR and vanilla NAR in Fig.~\ref{fig:err}). 
We believe that lack of context interdependence leads to increased substitution mistakes, particularly due to the conditional independence assumption required in single step NAR.
% We think the conditional independence assumption leads to the increased substitution mistakes of single step NAR.
\
Besides this, all of these NAR models were explored on academic benchmarks recorded from reading scenarios. Performance has not yet been assessed on a large scale industrial-level corpus.
\
This paper therefore aims to improve the single step NAR model so as that it can obtain recognition performance on par with an AR model on a large-scale corpus.

This work proposes a fast and accurate parallel transformer model~(termed \emph{Paraformer}) which addresses both challenges as stated above.
For the first, unlike previous CTC based works, we utilize a continuous integrate-and-fire~(CIF)~\cite{dong2020cif} based predictor net to estimate the target number and generate the hidden variables.
For the second challenge, we design a glancing language model~(GLM) based sampler module to strengthen the NAR decoder with the ability to model token inter-dependency. This is mainly inspired by work in neural machine translation~\cite{qian2020glancing}.
\
We additionally design a strategy to include negative samples, to improve performance by exploiting minimum word error rate~(MWER)~\cite{prabhavalkar2018minimum} training.

\
We evaluate \emph{Paraformer} on the public 178 hour AISHELL-1 and 1000 hour AISHELL-2 benchmarks, as well as an industrial 20,000 hour Mandarin speech recognition task.
\emph{Paraformer} obtains CERs of 5.2\% and 6.19\% on AISHELL-1 and AISHELL-2 respectively, which not only outperforms other recent published NAR models but is  comparable to the state-of-the-art AR transformer without an external language model.
As far as we know, \emph{Paraformer} is the first NAR model able to achieve comparable recognition accuracy to an AR transformer, and it does so with a 10x speedup on the large corpus.

\section{Methods}
\label{sec:Methods}

\subsection{Overview}
\label{overview}

\begin{figure}[t]
	\centering
	\includegraphics[width=0.8\linewidth]{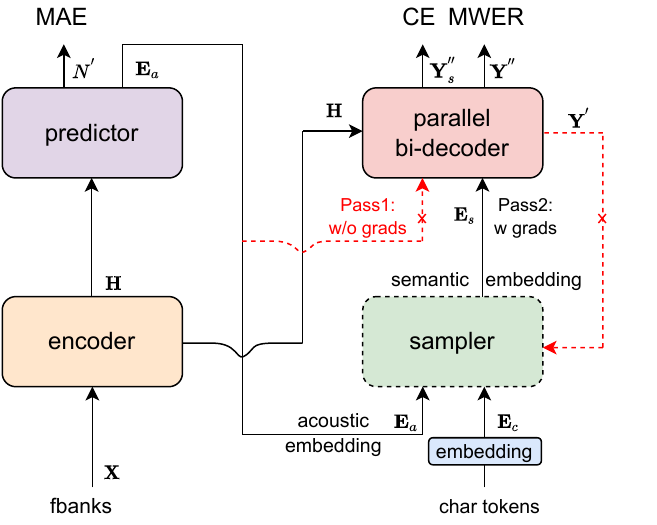}
	\caption{Structure of the proposed {Paraformer}.}
	\label{fig:nar}
	\vspace{-6mm}
\end{figure}

The overall framework of the proposed \emph{Paraformer} model is illustrated in Fig.~\ref{fig:nar}.
\
The architecture consists of five modules, namely the encoder, predictor, sampler, decoder and loss function.
\
The encoder is the same as an AR encoder, consisting of multiple blocks of memory equipped self-attention (SAN-M) and feed-forward networks (FFN)~\cite{gao2020san} or conformer~\cite{gulati2020conformer}.
\
The predictor is used to produce the acoustic embedding and guide the decoding.
The sampler module then generates a semantic embedding according to the acoustic embedding and char token embedding. 
The decoder is similar to an AR decoder except for being bidirectional. It consists of multiple blocks of SAN-M, FFN and cross multi-head attention~(MHA).
Besides the cross-entropy~(CE) loss, the mean absolute error~(MAE), which guides the predictor to convergence, and MWER loss, are combined to jointly train the system.

We denote the inputs as $(\mathbf{X}, \mathbf{Y})$, where $\mathbf{X}$ is the acoustic feature of frame number $T$, and $\mathbf{Y}$ is the target label with token number $N$.
The encoder maps  input sequence $\mathbf{X}$ to a sequence of hidden representations $\mathbf{H}$.
These hidden representations $\mathbf{H}$ are then fed up to the predictor to predict token number $N^{'}$ and produce acoustic embedding $\mathbf{E}_{a}$.
The decoder takes in acoustic embedding $\mathbf{E}_{a}$ and hidden representation $\mathbf{H}$ to generate target predictions $\mathbf{Y}^{'}$ for the first pass without backward gradients.
The sampler samples between acoustic embedding $\mathbf{E}_{a}$ and target embedding $\mathbf{E}_{c}$ to generate semantic embedding $\mathbf{E}_{s}$ according to the distance between predictions $\mathbf{Y}^{'}$ and target label $\mathbf{Y}$.
The decoder then takes in semantic embedding $\mathbf{E}_{s}$ as well as hidden representations $\mathbf{H}$ to generate final predictions $\mathbf{Y}^{''}$ for the second pass, this time with backward gradients.
Finally, the predictions $\mathbf{Y}^{''}$ are sampled to produce negative candidates for the MWER training, and the MAE is computed between target token number $N$ and predicted token number $N^{'}$.
Both MWER and MAE are jointly trained with a CE loss. %**IVM this was commented, but I think it's good to include (because the diagram shows one CE loss)

During inference, the sampler module is inactive and the bidirectional parallel decoder directly utilizes acoustic embeddings $\mathbf{E}_{a}$ and hidden representation $\mathbf{H}$ to output final prediction $\mathbf{Y}^{'}$ over only a single pass.
Although the decoder is operational in the forward direction twice during each \textit{training} stage, the computational complexity does not actually increase during \textit{inference} thanks to the single step decoding process.

\subsection{Predictor}
\label{predictor}

\begin{figure}[t]
	\centering
	\includegraphics[width=0.9\linewidth]{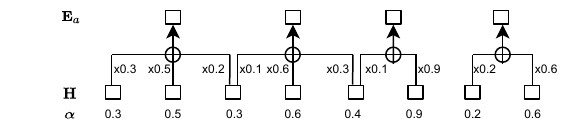}
	\caption{An illustration of the CIF process ($ \mathbf{\beta}$ is set to 1).}
	\label{fig:cif}
	\vspace{-6mm}
\end{figure}

The predictor consists of two convolution layers, with the output being a float weight $\alpha$ ranging from 0 to 1.
We accumulates the weight $\mathbf{\alpha}$ to predict token number.
An MAE loss $ \mathcal{L}_{MAE}=\left| \mathcal{N} - \sum ^{T}_{t=1} \alpha _{t} \right |  $ is added to guide the learning.
We introduce the mechanism of Continuous Integrate-and-Fire (CIF) to generate acoustic embedding.
CIF is a soft and monotonic alignment, which was proposed as a streaming solution for AED models in~\cite{dong2020cif}.
To generate acoustic embedding $\mathbf{E}_{a}$, CIF accumulates the weights $\mathbf{\alpha}$ and integrates hidden representations $\mathbf{H}$ until the accumulated weight reaches a given threshold $\mathbf{\beta}$, which indicates that an acoustic boundary has been reached ~(an illustration of this process is shown in Fig.~\ref{fig:cif}).
\
According to~\cite{dong2020cif}, the weight $\alpha$ is scaled by target length during training so as to match the number of acoustic embeddings $\mathbf{E}_{a}$ with target embeddings $\mathbf{E}_{c}$, while weight $\alpha$ is directly used to produce $\mathbf{E}_{a}$ for inference.
There may thus exist a mismatch between training and inference, causing the precision of the predictor to decay.
Since the NAR model is more sensitive to predictor accuracy than a streaming model, we propose using a dynamic threshold $\mathbf{\beta}$ instead of a predefined one to reduce the mismatch. The dynamic threshold mechanism is formulated as:

\begin{equation}
\mathbf{\beta}=\frac {\sum ^{T}_{t=1} \alpha _{t} } {\lceil \sum ^{T}_{t=1} \alpha _{t} \rceil}
\end{equation}

\subsection{Sampler}
\label{sampler}

In a vanilla single step NAR, the objective of its optimization could be formulated as:

\begin{equation}
\mathcal{L}_{\mathrm{NAT}}=\sum_{n=1}^{N} \log P\left(y_{n} \mid X ; \theta\right)
\end{equation}
\
However, as noted, the conditional independence assumption leads to inferior performance compared to an AR model.
Meanwhile the glancing language model (GLM) loss is defined as:

\begin{equation}
\mathcal{L}_{\mathrm{GLM}}=\sum_{y^{''}_{n} \in \overline{\mathbb{GLM}(Y, Y^{'})}} \log p\left[y^{''}_{n} \mid \mathbb{GLM}(Y, Y^{'}), X ; \theta\right]
\end{equation}
\
Where $\mathbb{GLM}(Y, Y^{'})$ denotes the subset of tokens selected by the sampler module between $\mathbf{E}_{c}$ and $\mathbf{E}_{a}$. And
$ \overline{\mathbb{GLM}(Y, Y^{'})} $ denotes the remaining unselected subset of tokens within the target $ \mathbf{Y} $. 
% The training loss above is calculated against these remaining token.
\

\begin{equation}
\mathbb{GLM}(Y, Y^{'})=\operatorname{Sampler}(\mathbf{E}_{s} \mid \mathbf{E}_{a}, \mathbf{E}_{c}, \lceil \lambda d(Y, Y^{'}) \rceil )
\end{equation}
\
Where $\lambda$ is a sampling factor to control the sample ratio. The $ d(Y, Y^{'}) $ term is the sampling number.   
It will be larger when the model is poorly trained, and should decrease along with the training process.
For this, we simply use the Hamming distance, defined as:
\
\begin{equation}
d(Y, Y^{'})= \sum_{n=1}^{N}\left(y_{n} \neq y^{'}_{n}\right)
\end{equation}
\

To summarize, the sampler module incorporates target embeddings $\mathbf{E}_{c}$ by randomly substituting $\lceil \lambda d(Y, Y^{'}) \rceil $ tokens into acoustic embedding $\mathbf{E}_{a}$ to generate semantic embedding $\mathbf{E}_{s}$. 
The parallel decoder is trained to predict the target tokens $ \overline{\mathbb{GLM}(Y, Y^{'})} $ with semantic context $\mathbb{GLM}(Y, Y^{'})$,  enabling the model to learn interdependency between output tokens.

\begin{table}[t]
	\centering
	%**IVM [new] I added ", without LM" in the caption below
	\caption{Comparison of ASR systems on AISHELL-1 and AISHELL-2 tasks (CER\%), without LM. $AR$ and $NAR$ denotes the AR baseline with beamsearch and proposed NAR method respectively reported in each work. ($\star$ : RTF is evaluated with batchsize of 8, $\dag$ : batchsize is unreported, batchsize of the others is 1).} 
%	\begin{tabular}[t]{@{\hspace{1.5pt}}c@{\hspace{1.5pt}}c@{\hspace{2.5pt}}|cc}
	\begin{tabular}[t]{lccl}
		\hline
		\textbf{AISHELL-1}& {AR / NAR}& {dev / test}& {RTF} \\\hline \hline

		\multirow{1}{*}{{A-FMLM}\cite{chen2019listen}}& NAR  & 6.2 / 6.7 & 0.2800 \\%\hline
% 		& NAR  & 6.2 / 6.7 & 0.2800 \\\hline

		\multirow{1}{*}{{Mask CTC}\cite{wang2021streaming}}& NAR  & 6.9 / 7.8 & 0.0500 \\%\hline

		\multirow{1}{*}{{LASO}~\cite{bai2021fast}}& NAR  & 5.9 / 6.6 & 0.0035 $\dag$ \\%\hline
% 		& NAR  & 6.2 / 7.0 & 0.0035 \\\hline

		\multirow{1}{*}{{NAT-UBD}\cite{zhang2021non}}& NAR &  5.1 / 5.6 & 0.0081 $\dag$ \\%\hline  
% 		{ }& NAR &  5.1 / 5.6 & 0.0081 \\\hline  
        {TSNAT}\cite{tian2021tsnat}& NAR &  5.1 / 5.6 & 0.0185 \\%\hline     		
        
		\multirow{2}{*}{{CTC-Enhanced}\cite{song2021non}}& AR  & 5.2 / 5.7 & 0.1703 \\
		& NAR &  5.3 / 5.9 & 0.0037 $\star$ \\%\hline    

		{{Improved }}& AR  & 4.8 / 5.2 & 0.2000 \\
		CASS-NAT\cite{fan2021improved}& NAR & 4.9 / 5.4 & 0.0230 \\\hline      		
		
% 		\multirow{3}{*}{\textbf{Ours}}& AR & 4.7 / 5.2 & 0.2100 \\
% 				& Vanilla-NAR & 4.7 / 5.3 & 0.0168 \\
% 		& \emph{Paraformer}  & \textbf{4.6} / \textbf{5.2} & 0.0168 \\\hline\hline
		
		\multirow{4}{*}{\textbf{Ours}}& AR & 4.7 / 5.2 & 0.2100 \\
				& Vanilla-NAR & 4.7 / 5.3 & 0.0168 \\
		& \multirow{2}{*}{\emph{Paraformer}}   & \multirow{2}{*}{\textbf{4.6} / \textbf{5.2}} & \textbf{0.0168} \\
		& & & \textbf{0.0026} $\star$ \\
		\hline\hline

		\textbf{AISHELL-2}& {AR / NAR}& {test\_ios}& {RTF} \\\hline \hline		
		
		\multirow{1}{*}{{LASO}~\cite{bai2021fast}}&  NAR  & 6.8  & - \\%\hline
		
		\multirow{2}{*}{{CTC-Enhanced}\cite{song2021non}}& AR  &  6.8 & 0.1703 \\
		 & NAR &  7.1  & 0.0037 $\star$ \\\hline

		\multirow{3}{*}{\textbf{Ours}}& AR & 6.18  & 0.2100 \\
		& Vanilla-NAR &  6.23 & 0.0168 \\
		& \emph{Paraformer}  & \textbf{6.19}  & 0.0168 \\
% 		& & & \textbf{0.0026} $\star$ \\
		\hline		
		
	\end{tabular}
	\label{tab:AISHLL1_state_of_art}
\end{table}

\subsection{Loss Function}
\label{loss}

There are three loss functions defined, namely the CE, MAE and MWER losses.
% The the MAE loss, the objective is to guide the predictor to convergence.
% And it could be defined as a regression task.
\
The types are jointly trained, as follows:

\begin{equation}
\mathcal{L}_{total}  = \gamma \mathcal{L}_{CE} + \mathcal{L}_{MAE} +\mathcal{L}_{\text {werr }}^{\text {N }}\left(\mathbf{x}, \mathbf{y}^{*}\right)
\end{equation}
\
For the MWER, it could be formulated as~\cite{prabhavalkar2018minimum}:

\begin{equation}\nonumber
\mathcal{L}_{\text {werr }}^{\text {N }}\left(\mathbf{x}, \mathbf{y}^{*}\right)=\sum_{\stackrel{  \mathbf{y}_{i} \in}{\operatorname{Sample}(\mathbf{x}, N)   }}   \widehat{P}\left(\mathbf{y}_{i} \mid \mathbf{x}\right)\left[\mathcal{W}\left(\mathbf{y}_{i}, \mathbf{y}^{*}\right)-\widehat{W}\right]
\end{equation}
%**IVM I tried to change the format slightly (no mathematical changes). Here is the original equation:
%\begin{equation}
%\mathcal{L}_{\text {werr }}^{\text {N }}\left(\mathbf{x}, \mathbf{y}^{*}\right)=\sum_{\mathbf{y}_{i} \in \operatorname{Sample}(\mathbf{x}, N)} \widehat{P}\left(\mathbf{y}_{i} \mid \mathbf{x}\right)\left[\mathcal{W}\left(\mathbf{y}_{i}, \mathbf{y}^{*}\right)-\widehat{W}\right]
%\end{equation}
\
There is only one output path for NAR models due to the greedy search decoding. 
As noted above, we exploit the negative sampling strategy to generate multiple candidate paths by randomly masking the top1 score token during the MWER training.

\begin{table}[t!]
	\centering
	
	\caption{Evaluation of sampling ratio (CER\%).}
	\begin{tabular}[t]{cccccc}
		\hline
		$\lambda$&   0.2 &        0.5 & 0.75 & 1.0 & 1.5 \\\hline%\hline
		Far-field & 14.64& 	14.37&  \textbf{14.17}&	14.24& 	14.47 \\
		Common & 8.22& 	8.13& \textbf{7.98}&	8.09& 	8.13   \\ \hline
		% Live-field & 17.83&  17.54&	\textbf{17.32}& 17.55& 17.75 \\ \hline
		
	\end{tabular}
	\label{tab:weight}
\end{table}

\section{Experiments}
\label{sec:exp}

\subsection{Experimental Setup}
We evaluated the proposed methods on the openly available AISHELL-1 (178-hours)~\cite{bu2017aishell}, AISHELL-2 (1000-hours) benchmarks~\cite{du2018aishell}, plus a 20,000 hour industrial Mandarin task. 
\
% For AISHELL-2, we train using all 1000 hours of training data, reserving $dev\_ios$ and $test\_ios$ sets for validation and evaluation, respectively. 
\
The latter task is the same large corpus as in~\cite{gao2020san,gao2020universal}. %, and consists of multi-domain recordings from genres including news, sports, tourism, games, literature, education and so on. 
% This is divided into training and development sets by a ratio of $95:5$. 
A \emph{far-field} set  of about 15 hours data and a \emph{common} set  of about 30 hours data are used to evaluated the performance.
Other configurations could be found in~\cite{gao2020san,gao2020universal,zhang2020streaming}. 
\
Real time factor (RTF) was used to measure the inference speed on GPU (NVIDIA Tesla V100). %, with batchsize of 1.
% Acoustic features used in all experiments are 80-dimensional log-mel filter-bank (FBK) energies computed over 25ms windows with 10ms shift.
% We stack consecutive frames within a context window of 7 (3+1+3) to produce 560-dimensional features and then down-sample the input frame rate to 60ms. 
% Acoustic modeling units are Chinese characters, totalling 5211 and 9000 for AISHELL-2 and the 20,000 hour tasks respectively. 
% Models are trained with Tensorflow in a distributed manner. %~\cite{abadi2016tensorflow}. %with 2 GPUs and 32 GPUs for AISHELL-1 and 20000-hour task, respectively.
% As to the detailed experimental setup,  we adopt LazyAdamOptimizer with $\beta_1=0.9$, $\beta_2=0.999$, and the strategy for learning rate is \emph{noam\_decay\_v2} with $d_{model}=512, warmup\_n=8000, k=4$. 
% Label smoothing and dropout regularization with a value of $0.1$ are incorporated to prevent over-fitting. 
% SpecAugment~\cite{park2019specaugment} is also used in all experiments.
Our code is available in FunASR \footnote{https://github.com/alibaba-damo-academy/FunASR}.

\begin{table*}[t]
	\centering
	
	\caption{Performance of three systems on the industrial 20,000 hour task (CER\%).}
	\begin{tabular}[t]{c|c|ccccc|ccc}
		\hline
		\multirow{1}{*}{{\textbf{Parameter}}}& \multirow{1}{*}{\textbf{-}}& \multicolumn{5}{c|}{\textbf{Transformer-SAN-M (41M)}}& \multicolumn{3}{c}{\textbf{Transformer-SAN-M-large (63M)}} \\\cline{1-10}		
		Model & CTC &    AR &       Vanilla NAR & \multicolumn{3}{|c|}{\emph{Paraformer}} & AR &       Vanilla NAR & \emph{Paraformer}  \\\hline
		Dynamic $\mathbf{\beta}$ & -& -& w& w/o& w& w& -& w& w \\
		MWER & -& -& -& w/o& w/o& w& -& -& w \\\hline\hline
		RTF & - &  0.067 & 	0.007 &  0.007 & 0.007 & 0.007	 & 	    0.094& 	 0.009& 	 0.009 	 \\ \hline
		Far-field& 17.71&  13.76& 	16.39 & 14.39 & 14.17& 14.07& 	   12.57& 	  14.86&    12.93	 \\
		Common & 9.93 &7.75& 	9.35&  8.15& 7.98&	7.86& 	    7.55& 	 8.67& 	 7.71 	 \\ \hline
		% Live-field & 24.35& 17.20& 	19.52& 17.54& 17.32&	17.30& 	    15.43& 	 17.66& 	16.06  	 \\ \hline

	\end{tabular}
	\label{tab:20000hour}
\end{table*}

\subsection{AISHELL-1 and AISHELL-2 task}
\label{sec:AISHELL-2}
The AISHELL-1 and AISHELL-2 evaluation results are detailed in Table~\ref{tab:AISHLL1_state_of_art}. 
    For fair comparison with published works, RTF is evaluated on ESPNET~\cite{watanabe2018espnet}.
    No external language model (LM) nor unsupervised pre-training is used with any of the experiments in Table~\ref{tab:AISHLL1_state_of_art}.
    \
    For the AISHELL-1 task, we firstly trained an AR transformer as baseline, with the configuration matching the AR baseline in~\cite{fan2021improved}. The performance of the baseline is state-of-the-art among AR transformers, excluding system with large scale knowledge transfer such as~\cite{deng2022_knowledge_transfer} since we aim for architectural improvement rather than gains from a bigger dataset. %**IVM [new] suggested sentence
    \
    The vanilla NAR shares the same structure with our proposed model \emph{Paraformer}, but without the sampler. %**IVM can we add this ----> and two-step training. 
    % From Table~\ref{tab:AISHLL1_state_of_art}, 
    It can be seen that our vanilla NAR surpasses the performance of other recent published NAR works, $e.g.$, improved CASS-NAT~\cite{fan2021improved} and CTC-enhanced NAR~\cite{song2021non}.
    \
    Nevertheless, its performance is slightly inferior to the AR baseline due to the lack of context dependency between output tokens.
    But when we enhance the vanilla NAR with GLM via the sampler module in \emph{Paraformer}, we obtain comparable performance to the AR model.
    \
    While \emph{Paraformer} obtains a recognition CER of 4.6\% and 5.2\% on dev and test set respectively, the inference speed (RTF) is more than 12 times faster than the AR baseline.
    % Although we test without a LM, we confirm that if one is used, test set CER reduces further to 4.8\%. %**IVM [new] suggested sentence
    \
    For the AISHELL-2 task, the model configuration is the same with AISHELL-1.
    From Table~\ref{tab:AISHLL1_state_of_art}, it can be seen that the performance gains are similar to those for AISHELL-1. 
    \
    Specifically, \emph{Paraformer} achieved a CER of 6.19\% for the $test\_ios$ task, with more than 12 times faster inference speed.
    \
    As far as the authors are aware, this is the state-of-the-art performance among NAR models on both AISHELL-1 and AISHELL-2 tasks.

\subsection{Industrial 20,000 hour task}
\label{sec:20000}

Extensive experiments were used to evaluate our proposed approach, detailed in Table~\ref{tab:20000hour}. 
% The normal and large models refer to the dimensions of SAN-M and FFN being either 256-1024 and 320-1280 respectively, as detailed in~\cite{gao2020san}.
Dynamic $\mathbf{\beta}$ denotes the dynamic threshold detailed in Section~\ref{predictor} while
CTC refers to the DFSMN-CTC-sMBR system with LM~\cite{zhang2019investigation}.
RTF is evaluated on OpenNMT~\cite{klein-etal-2017-opennmt}.

Looking first at the models with a size of 41M,
the AR baseline with attention dimension of 256 is the same as in~\cite{gao2020san}.
We can see a different phenomenon to that noted in the Sec.~\ref{sec:AISHELL-2}. Here we find that the CER of vanilla NAR differs from that of the AR model by a large margin.
% We believe the reason is the large scale industrial-level data is more complicated than academic benchmarks.
Nevertheless, vanilla NAR still outperforms CTC, which makes a similar conditional independence assumption.
\
When equipped with GLM, \emph{Paraformer} obtains 13.5\% and 14.6\% relative improvement on \emph{Far-field} and \emph{Common} tasks respectively, compared to vanilla NAR.
When we further add the MWER training, the accuracy improves slightly.
More importantly, \emph{Paraformer} achieves comparable performance to the AR model (less than 2\% relative loss) with the benefit of 10x faster inference speed.
\
We have also evaluated the dynamic threshold for CIF. From Table~\ref{tab:20000hour}, it is evident that the dynamic threshold helps  to further improve accuracy.
Compared to a predefined threshold as in CIF, the dynamic threshold reduces the mismatch between inference and training, to extract acoustic embedding more accurately.%predict token number more accurately.

\
Evaluating on the larger model size of 63M, the phenomenon seen is similar to that noted above. 
Here, \emph{Paraformer} achieves 13.0\% and 11.1\% relative improvement on \emph{Far-field} and \emph{Common} task respectively, over vanilla NAR.
Again, \emph{Paraformer} achieves comparable accuracy to the AR model (less than 2.8\% relative difference), again achieving 10x speedup.
If we compare \emph{Paraformer-63M} against AR transformer-41M, although the \emph{Paraformer} model size is larger, its inference speed improves (RTF from 0.067 to 0.009). Hence \emph{Paraformer-63M} can achieve a 6.0\% relative improvement over the AR transformer-41M on the \emph{Far-field} task, while accelerating the inference speed by 7.4 times.
This reveals that \emph{Paraformer} can achieve superior performance through increased model size, while still maintaining a faster inference speed than AR transformer.

Finally, we evaluate the sampling factor $\lambda$ in the sampler, as shown in Table~\ref{tab:weight}.
As expected, the recognition accuracy improves as $\lambda$ increases, due to the better context provided by the targets.
However when the sampling factor is too large, it will cause a mismatch between training and inference, where we decode twice with targets for training and decode once without targets for inference. 
Nevertheless, the performance of \emph{Paraformer} is robust to $\lambda$ in a range from 0.5 to 1.0.
% In particular, the sampling factor of 0.75 seems slightly better.

\subsection{Discussion}
\label{sec:disscussion}

From the above experiments, we note that, compared to the AR model, the performance decay of vanilla NAR on AISHELL-1 and AISHELL-2 task is small, but is much bigger for the large scale industrial-level corpus.
Compared to academic benchmarks from a reading corpus (e.g. AISHELL-1 and -2) , the industrial-level dataset  reflects more complicated scenarios, and thus is more reliable for evaluating NAR models.
As far as we know, this is the first work which explores NAR models on a large-scale industrial-level corpus task.

The experiments above show that \emph{Paraformer} obtains significant improvement compared to vanilla NAR by over 11\% while \emph{Paraformer} performs similarly to well-trained AR transformer.

To understand why, we performed further analysis.
First, we determined the error type statistics for the AR, vanilla NAR and \emph{Paraformer} models on the 20,000 hour task, plotted in Fig.~\ref{fig:err}.
We counted the total number of errors types, namely insertion, deletion and substitution respectively on \emph{Far-field} and \emph{Common}, and normalized them by the total number of target tokens.
The vertical axis of Fig.~\ref{fig:err} is the ratio of error types.
\
We can see that, compared to the AR system performance, the insertion errors in the vanilla NAR increase slightly, while deletion decreases to a small extent.
This indicates that the accuracy of the predictor is superior, with the help of the dynamic threshold.
However, the substitution errors rise dramatically, which explains the large gap in performance between them. 
We believe this is caused by the conditional independence assumption within the vanilla NAR model.
Compared to vanilla NAR, the substitution errors in \emph{Paraformer} decrease significantly, accounting for most of its performance improvement.
We believe the decline in substitution is because the enhanced GLM enables the NAR model to better learn inter-dependency between output tokens.
\
Nevertheless, compared to AR, there still exists a small gap in the number of substitution errors, leading to the slight difference in recognition accuracy.
We think the reason is that the beam search decoding of AR could play a strong role in the language model compared to the glancing language model.
To eliminate this remaining performance gap, we aim to combine \emph{Paraformer} with an external language model in our future work.

\section{Conclusion}
\label{sec:conclusion}

This paper has proposed a single-step NAR model, \emph{Paraformer}, to improve the performance of NAR end-to-end ASR systems.
We first utilize a continuous integrate-and-fire (CIF) based predictor to predict the token number and generate hidden variables.
We improved CIF with a dynamic threshold instead of a predefined one, to reduce the mismatch between inference and training.
Then we design a glancing language model based sampler module to generate semantic embeddings to enhance the NAR decoder's ability to model the context interdependence.
Finally, we designed a strategy to generate negative samples in order to perform minimum word error rate training to further improve performance.
\
Experiments  conducted on the public AISHELL-1 (178 hours) and AISHELL-2 (1000 hours) benchmark as well as an industrial-level 20,000 hour corpus show that the proposed Paraformer model can achieve comparable performance to the
state-of-the-art AR transformer with 10x speedup. %well-trained AR models on three tasks.
% We have also made extensive experiments to evaluate the effectiveness of the three efforts.
% In future work, we would investigate to integrate language model into \emph{Paraformer}.

\vfill\pagebreak

\bibliographystyle{IEEEtran}

\bibliography{mybib}

% \begin{thebibliography}{9}
% \bibitem[1]{Davis80-COP}
%   S.\ B.\ Davis and P.\ Mermelstein,
%   ``Comparison of parametric representation for monosyllabic word recognition in continuously spoken sentences,''
%   \textit{IEEE Transactions on Acoustics, Speech and Signal Processing}, vol.~28, no.~4, pp.~357--366, 1980.
% \bibitem[2]{Rabiner89-ATO}
%   L.\ R.\ Rabiner,
%   ``A tutorial on hidden Markov models and selected applications in speech recognition,''
%   \textit{Proceedings of the IEEE}, vol.~77, no.~2, pp.~257-286, 1989.
% \bibitem[3]{Hastie09-TEO}
%   T.\ Hastie, R.\ Tibshirani, and J.\ Friedman,
%   \textit{The Elements of Statistical Learning -- Data Mining, Inference, and Prediction}.
%   New York: Springer, 2009.
% \bibitem[4]{YourName17-XXX}
%   F.\ Lastname1, F.\ Lastname2, and F.\ Lastname3,
%   ``Title of your INTERSPEECH 2022 publication,''
%   in \textit{Interspeech 2022 -- 23\textsuperscript{rd} Annual Conference of the International Speech Communication Association, September 18-22, Incheon, Korea, Proceedings, Proceedings}, 2022, pp.~100--104.
% \end{thebibliography}

\end{document}